\documentclass[sigconf]{acmart}
\renewcommand\footnotetextcopyrightpermission[1]{} % removes footnote with conference information in first column
\usepackage{tabularx,booktabs,caption}
\usepackage{array,multirow}
\usepackage{xcolor}
\definecolor{newgray}{RGB}{196,196,196}
\definecolor{LightGray}{gray}{0.9}
\usepackage{colortbl}
\usepackage{booktabs}
\usepackage{threeparttable}

\setcopyright{none}
\begin{document}

\title{Towards Better Driver Safety: Empowering Personal Navigation Technologies with Road Safety Awareness}


\author{Runsheng Xu}
\affiliation{%
  \institution{University of California, Los Angeles}}
\email{rxx3386@ucla.edu}

\author{Shibo Zhang}
\affiliation{%
  \institution{Northwestern University}
}
\email{shibo.zhang@northwestern.edu}

\author{Yue Zhao}
\affiliation{%
  \institution{Carnegie Mellon University}
  \city{Pittsburgh, PA}
  \country{United States}}
\email{zhaoy@cmu.edu}

\author{Peixi Xiong}
\affiliation{%
  \institution{Northwestern University}
}
\email{peixixiong2018@u.northwestern.edu}

\author{Allen Yilun Lin}
\affiliation{%
  \institution{Northwestern University}
}
\email{allen.lin@eecs.northwestern.edu}

\author{Brent Hecht}
\affiliation{%
  \institution{Northwestern University}
}
\email{bhecht@northwestern.edu}
\authornote{Corresponding authors}

\author{Jiaqi Ma }
\affiliation{%
  \institution{University of California, Los Angeles}
}
\email{jiaqima@ucal.edu}
\authornotemark[1]

\renewcommand\shortauthors{Xu et al.}

\begin{abstract}
Recent research has found that navigation systems usually assume that all roads are equally safe, directing drivers to dangerous routes, which led to catastrophic consequences. To address this problem, this paper aims to begin the process of adding road safety awareness to navigation systems. To do so, we first created a definition for road safety that navigation systems can easily understand by adapting well-established safety standards from transportation studies. Based on this road safety definition, we then developed a machine learning-based road safety classifier that predicts the safety level for road segments using a diverse feature set constructed only from publicly available geographic data. Evaluations in four different countries show that our road safety classifier achieves satisfactory performance. Finally, we discuss the factors to consider when extending our road safety classifier to other regions and potential new safety designs enabled by our road safety predictions.
\end{abstract}



%\ccsdesc[500]{Information interfaces and presentation (e.g., HCI)}


\maketitle
% `=============================================
\pagestyle{plain}
% =============================================`

% Introduction
\section{Introduction}
While reliable in most cases, personal navigation technologies (e.g., Google Maps) occasionally route drivers to dangerous roads, resulting in catastrophic incidents \cite{casner2016challenges}. Prior HCI research identified 47 catastrophic incidents in which navigation systems directed drivers to unpaved roads, narrow lanes, and roads with low clearance, causing drivers to get stranded, to crash with roadside objects, and to hit overpasses \cite{lin2017understanding}. In these incidents, drivers bore financial losses, suffered severe injuries, and, in extreme cases, lost their lives.

\begin{figure}
\centering
\includegraphics[width=1\columnwidth]{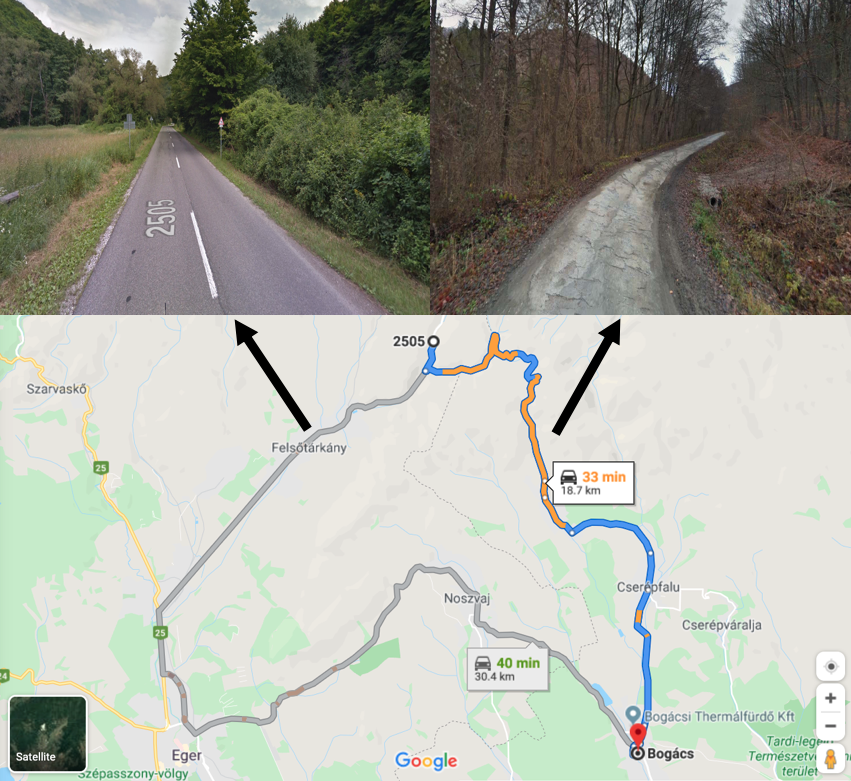}
\caption{Without road safety information, Google Maps recommended a faster but more dangerous route (highlighted). By contrast, the alternative route (grey) is slower but safer. Images are from Google Maps and Google StreetView.}
\label{fig:compare}
\end{figure}

One key reason behind these catastrophic incidents is that personal navigation systems usually lack road safety information and thus assume all roads are equally safe \cite{lin2017understanding}. As such, navigation systems might generate routes that prioritize other parameters such as travel time over travel safety, resulting in short-yet-dangerous routes. Indeed, many short-yet-dangerous route recommendations that potentially lead to crashes can be found on Google Maps, the most popular personal navigation system used by over 50 million people every day \cite{googlemaps}. For instance, Figure \ref{fig:compare} compares two routes with the same origins and destinations generated by Google Maps in a rural area of Hungary. Without road safety information, Google Maps recommended a route that is faster but more dangerous (highlighted). The roads on this route are narrow and poorly surfaced, which, according to transportation research \cite{beenstock2000globalization}, are more dangerous for drivers. In comparison, the alternative route (gray) is slower but safer - it passes through roads that are wide and well-maintained.

To help navigation systems generate safer routes for drivers, we sought to begin the process of adding road safety awareness into personal navigation devices. We accomplished this goal in two stages. First, we developed a straightforward road safety definition that personal navigation systems can easily understand by adapting well-established safety standards from transportation studies. Specifically, we focused on the Star Ratings standard from International Road Assessment Programme (iRAP), which gives a single measure of the "built-in" safety for every 100-meter road segment (i.e., safety measured by a road design alone) \cite{iRAPStar}. Based on the Star Ratings standard, we created a straightforward binary definition for high-risk roads and safe roads, which navigation systems can smoothly incorporate.

In the second stage of our work, we utilized machine learning to produce large-scale predictions using our binary road safety definition. Although the road safety labels we defined can be directly converted from the Star Ratings scores, existing Star Ratings scores only cover a small percentage of roads, which is insufficient for navigation systems. To increase the coverage of road safety labels, we developed a classifier that automatically predicts high-risk roads and safe roads. Our classifier relied on a key insight: many road attributes that are highly relevant to road safety are with open access and can be used as features to predict the road safety labels. Leveraging open data from geographic peer-production projects and government transportation agencies as features, we built and tested our road safety classifier with straightforward machine learning algorithms in four different countries (Greece, Slovenia, Croatia, and the Netherlands). Our results show that our model achieved useful performance, consistently outperforming baseline approaches.

We continue below with an overview of related work, followed by a description of our road safety definition in the context of navigation systems. We then describe the construction and evaluation of our road safety classifier, which predicts the road safety labels according to our definition. Finally, we discuss how our road safety classifier may be extended to regions beyond the four countries studied in this paper, and how road safety awareness can enable new technology designs that significantly enhance the safety of drivers who use navigation systems.

% Related Work
\section{Related Work}
\subsection{Risks Associated with Navigation Systems}

Many HCI studies have investigated the risks associated with personal navigation devices. While early work primarily focused on identifying (e.g., \cite{kun2009glancing,bach2009interacting,jensen2010studying}) and addressing (e.g., \cite{tarqui2013reducing,bark2014personal,voice_interact}) risks related to distraction introduced by the navigation systems, recent studies have highlighted other types of risks, including the risks of recommending unsafe routes. Most notably, Lin et al. \cite{lin2017understanding} analyzed 158 catastrophic incidents associated with personal navigation systems and found that 47, or about 1/3 of the events they collected, were attributed to dangerous routes recommended by the navigation systems. For example, Lin and colleagues found that personal navigation systems sometimes routed drivers through roads that are unpaved, narrow, or with low clearance, causing vehicles to get stranded, to crash with roadside objects, or to hit overpasses. Furthermore, Lin et al. also highlighted a set of road attributes, including road surface, lane width, and vertical clearance, that are critical for navigation systems to consider in order to generate safe routes for drivers.

Our paper has a twofold relationship with the work by Lin et al. First, our work directly addressed the risks associated with dangerous routes identified in their work by adding road safety awareness into navigation systems. Second, as we show later, the road attributes that Lin et al. recognized as critical to road safety informed how we defined and predicted road safety for navigation systems.

\subsection{Safe Routes for Driving}
Previous navigation and routing research have used the term "safe routes" in very different contexts. Among these cases, researchers primarily used "safe routes" to refer the routes that avoid high-crime areas defined by public safety data and sentiments from crowdsourced data (e.g., \cite{CrimeAle32:online,shah2011crowdsafe,kim2014socroutes,fu2014treads,galbrun2016urban}). Other researchers have used this term to refer to routes that avoid natural hazards, such as inundations \cite{kim2010proposal} and hurricanes \cite{li2014routing}. While these studies focused on different aspects of "safety", their use of a combination of government and crowdsourced data for defining the safety score inspired our solution. 

More similar to our work, a few studies used "safe routes" to refer to the safety defined on the road physical attributes \cite{kortge2006system}. For example, Kortge and Zhang \cite{kortge2006system} developed a navigation system and algorithm that will optimize routes for safe road characteristics such as lane width, the number of lanes, and light condition. However, a key constraint for such systems is that they can only operate in areas where road safety information is available. As we will show later, reliable road safety information is extremely dearth, significantly limiting the applicability of these systems. By comparison, our work does not focus on developing a safety-optimized routing algorithm but focuses on how to define and predict road safety data which can be later used in these safety-optimized routing algorithms. 

To avoid confusion on terminology, we clarify that unless otherwise specified, the rest of the paper will use \textit{safety} to exclusively refer to the safety associated with road design. 

\subsection{Machine Learning Applications}
The recent advancements in machine learning have significantly prompt the developments of many fields, such as autonomous driving~\cite{xu2021opencda,xu2021opv2v,xu2021holistic, xia2018automated,xia2022estimation, xiong2019imu, liu2018intelligent, liu2020vision,liu2021automated, paz2020probabilistic, christensen2021autonomous}, intelligent transportation system~\cite{li2021domain, fu2021let, zhao2020fusion}, medical system~\cite{du2020interpretable,du2021disease,du2021graphgt,rahman2021generative,liu2021deep} and image processing~\cite{tu2021temporal, tu2021ugc, tu2021video, chen2020proxiqa}. Among these machine learning algorithms, XGBoost\cite{xgboostfirst} has been popular and effective in a variety of applied machine learning tasks (e.g. \cite{hengl2017soilgrids250m,chen2018rise,segler2018generating}) because of its scalable and accurate implementation of gradient boosting machines. In this paper, we target to employ XGBoost on a new topic -- road safety prediction.

% Safety definition
\section{Defining Road Safety for Navigation Systems}
In this section, we detail how we defined road safety for navigation purposes. We first describe how we selected an appropriate safety standard from transportation studies and then explain how we simplified this standard to create a straightforward definition that navigation systems can easily understand.

Road safety standards developed by transportation researchers can be broadly classified into two types. The first type of standard is based on crash statistics (e.g., \cite{road_safty1,road_safty7,road_define_1}), which computes a road safety score as the ratio of the number of crashes to a certain measure of exposure such as the population in a region. The second type of standard is road attribute-based (e.g., \cite{Consurement2,iRAPStar}) and computes the road safety score as a polynomial of many design attributes of the roads, including but not limited to width, surface, and vertical clearance.

We selected the most widely used road attribute-based standard - the Star Ratings standard \cite{iRAPStar} - as the basis for our road safety definition. Unlike the crash statistics-based standards which are influenced by factors irrelevant to road's physical safety such as local driving habit \cite{drive_habit}, road attributed-based standards, especially Star Ratings, measure only the safety that is "built-in" to the road \cite{iRAPStar}. Proposed and maintained by the International Road Assessment Programme (iRAP) \footnote{https://www.irap.org/}, the Star Ratings standard has been partially implemented in more than 100 countries around the world. The Star Ratings standard calculates a single score ranging from 1-star (most dangerous) to 5-star (safest) using a formula that involves an extensive list of safety-related road design attributes. The Star Ratings standard is particularly attractive to our purpose because its definition incorporates all road attributes that Lin et al. identified as the major contributors to GPS-related catastrophic incidents \cite{lin2017understanding}, such as width, surface, and vertical clearance of the roads.

However, Star Ratings standard can not be directly used by navigation systems because the five-star systems are difficult to interpret for navigation purposes. As a standard created for transportation studies, one of the main purposes of the Star Ratings is to offer detailed and accurate information to evaluate the cost of road improvement projects \cite{irapsaveplan}. As such, while the five-star system in the Star Ratings is perfect for differentiating the amounts of money required for the road improvement projects, it does not offer interpretations about which roads are desirable for drivers to travel - a critical piece of information for navigation systems.

To address the lack of interpretability issues, we simplified the Star Ratings standard to create our road safety definition. Among the five levels in the Star Ratings system, the three-star level is a critical threshold. Most notably, three-star is considered as an economical criterion to ensure a desirable safety level by iRAP \cite{iRAPStar} and has been adopted by the World Bank and other government agencies as a requirement for the roads construction projects \cite{congress2015} around the world. In other words, roads with three stars and above are considered to have reached the minimum safety for traveling purposes.

Leveraging this critical threshold, we collapsed the five-star system in the Star Ratings into two classes and created the following binary definitions for road safety: 
\begin{itemize}
    \item \textbf{Safe roads}: road segments with three stars and above
    \item \textbf{High-risk roads}: road segments with one or two stars.
\end{itemize}

%Prediction
\section{Predicting Road Safety}

Using the above binary definition of road safety, it is trivial to convert the existing Star Ratings data into road safety labels. However, the resulting road safety labels have one significant limitation - they do not cover road segments without existing Star Rating scores. In fact, due to the high implementation cost of the Star Ratings standard which requires manually labeling road design attributes from images of the road, the existing Star Ratings scores are extremely sparse. For example, according to iRAP's official statistics \cite{iRAPStar}, even in regions that have the best implementation such as Europe, only an average of 5.4\% of road segments are rated with Star Ratings scores. As such, any binary road safety labels generated from these Star Ratings scores do not have sufficient coverage for personal navigation technologies.

To scale up the coverage of our road safety labels, we proposed and evaluated a machine learning-based road safety classifier that automatically predicts our binary road safety labels for road segments. As such, the structure of this section follows the best practices of applied machine learning research  (e.g., \cite{hciconfai_1, hciconfai_2, crimepredict}). We first define the feature set used in our road safety classifier. We then summarize how we collected the ground truth datasets, including both the features and the road safety labels, and highlight the unique characteristics of these data. Finally, we detail the training and evaluation strategies and find that our classifier achieves useful performances that outperform the baseline approaches.

\subsection{Feature Definitions}
Our road safety classifier is only as powerful as the features it leverages. Our choice of features is grounded in well-established theories from transportation studies that demonstrated these features' high relevance to road safety. In addition, to facilitate other researchers to replicate our results and leverage our approach, we purposefully only selected features that come from two publicly accessible sources - geographic peer-production projects and government transportation agencies.

\subsubsection{OpenStreetMap Features}
The most important features used in our road safety classifier are the features from OpenStreetMap (OSM). As known as the "Wikipedia of maps" \cite{osmwikimaps}, OpenStreetMap is a highly successful geographic peer-production project in which volunteers gather to produce free and accurate data for all geographic entities, including roads, for the entire world. Geographic data from OSM have been widely used in a variety of commercial-level map products such as Bing Maps and Facebook Live Maps \cite{osmconsumer}.

The 12 OSM features we selected, shown in the top section of Table \ref{table:features_definition}, can be broadly classified into two categories. The first category includes the OSM data that directly map to road attributes used in the Star Ratings standard. Because our road safety label is defined based on the Star Rating standard, we hypothesized that these OSM features would be highly relevant to our road safety labels. We manually checked the definitions of over 100 road attributes defined by OSM community \cite{MapFeatu67:online} and identified eight such OSM features such as maximum speed, road surface, and road lighting condition.

The second category of OSM features that our road safety classifier used characterizes the accessibility from the location of a road segment. Accessibility in transportation planning describes how easy it is to reach other areas from a given location \cite{litman2003measuring}. We selected features about accessibility because prior research in transportation studies has established that places with lower accessibility are associated with poorer road safety \cite{handy2001evaluating,antunes2003accessibility,olsson2009improved}. In this paper, we focused on accessibility metrics that solely depend on roads' topological structure. Specifically, we used isochrones, which show the area a person can reach from a given location within a given period \cite{o2000using}. We computed the 15-minute isochrone from the midpoint of the road segment and included the area of the isochrone, the total population within the isochrone, and the reach factor (detailed in Table \ref{table:features_definition}) as our features.

\subsubsection{Traffic Statistics Features}
In addition to features from OSM, we also included traffic statistics as features. We do so because prior research in transportation studies found that traffic statistics, such as the operating speed and traffic volume, are highly correlated to road safety \cite{golob2003relationships}. The bottom section of Table \ref{table:features_definition} details the traffic statistics features, including annual average daily traffic, the mean operating speed of cars, the 85\% operating speed of cars.

However, compared with the OSM features, traffic statistics features suffered from a major drawback. While OSM data around the world are accessible through a centralized repository maintained by the OSM community \footnote{https://wiki.openstreetmap.org/wiki/Downloading\_data}, the accessibility of traffic statistics highly varies - local and national government agencies make own decision on whether to publish the data or not.

We took this ecological validity concern into account when designing our machine learning experiments. Specifically, we trained and evaluated two versions of our classifier - one with all features and the other with OSM features only - in order to gain a comprehensive understanding of the ecological validity of our model. As we show later, parsimonious models with only OSM features performed almost as good as the models trained on all features, highlighting that our model is robust in areas where traffic statistics are unavailable.

% \usepackage{multirow}

% \begin{table}[h!]
\begin{table*}[h!]

\caption{Features for our road safety classifier}
\begin{tabular}{p{2.2cm}p{2.5cm}p{12cm}}
\hline
\rowcolor{LightGray} \textbf{Source} & \textbf{Feature Name}       & \textbf{Definition}                                                                              \\ \hline
\multirow{12}{3.3cm}{OSM Features}               & highway            & one-hot encoding indicating the type of the road                                        \\ 
                                            % \cline{2-3} 
                                             & oneway             & a binary value of whether this road segment is one way                                  \\ 
                                            %  \cline{2-3} 
                                             & maxspeed           & an integer for the maximum speed of this segment                                        \\ 
                                            %  \cline{2-3} 
                                             & surface            & one-hot encoding of the surface material                                                \\ 
                                            %  \cline{2-3} 
                                             & smoothness         & an integer for road surface quality level                                               \\ 
                                            %  \cline{2-3} 
                                             & lit                & a binary value of whether street lights equipped                                        \\ 
                                            %  \cline{2-3} 
                                             & bridge             & a binary value of whether there is a bridge above or below this road segment            \\ 
                                            %  \cline{2-3} 
                                             & lanes              & an integer for the number of lanes                                                      \\ 
                                            %  \cline{2-3} 
                                             & nodes              & an integer for the number of nodes nearby the road segment                     \\ 
                                            %  \cline{2-3} 
                                             & iso area           & a float value of the area of isochrone                                                  \\ 
                                            %  \cline{2-3} 
                                             & reach factor       & \begin{tabular}[c]{@{}l@{}}a float value that measures the complexity of the road network around this road segment \cite{orsgithub}\end{tabular} \\ 
                                            %  \cline{2-3}
                                             & totpop             & an integer for the population covered by the isochrone                                  \\ 
                                             \hline
\multirow{4}{3.3cm}{Traffic Statistics \\Features} & direction          & the driving direction of this segment                                                   \\ 
                                            % \cline{2-3} 
                                             & aadt               & an integer for the annual average daily traffic of the road segment                     \\ 
                                            %  \cline{2-3} 
                                             & mean speed of cars & an integer for vehicle mean operating speed                                             \\ 
                                            %  \cline{2-3} 
                                             & 85\% speed of cars & an integer for the vehicle 85th percentile                                              \\ 
                                             \hline
\end{tabular}
\label{table:features_definition}

% \caption{Results}\label{tab:table}
% \setlength\tabcolsep{3.5pt} % default value: 6pt
% \begin{tabularx}{\columnwidth}{@{} Z *{6}{c} @{}}
% \toprule 
% Name & \multicolumn{3}{c}{Kannada} & \multicolumn{3}{c@{}}{Malayalam}\\
% % \cmidrule(lr){2-4} \cmidrule(l){5-7} 
% % & (P)\% & (R)\% & (F)\% & A & B & C\\
% \midrule 
% Morfessor CAP     &  48.07&60.39    &    53.53 & 47.25 &60.01  &52.88   \\ 
% Morpheme induction& 66.78 & 57.97  &  62.07  &60.33   & 59.55  &59.95  \\
% \bottomrule 
% \end{tabularx}
\end{table*}

\subsection{Ground Truth Datasets}
\subsubsection{Building Ground Truth Datasets}
We prepared four ground truth datasets, each representing a sample of road segments from the following four countries - Croatia, Slovenia, Greece, and the Netherlands. As we show later, these four countries were selected because they represent countries with relatively different OSM feature qualities. By conducting experiments on ground truth datasets with varying feature qualities, we gain a deeper understanding of how well our road safety classifier will generalize to other regions.

To build these four ground truth datasets, we first generated binary road safety labels. To do so, we downloaded the existing Star Ratings data in these four countries, including both the coordinates of the road segments and the Star Rating scores of these road segments, from iRAP website \cite{iRAPStar}. We then converted these Star Rating scores to binary road safety labels.

For each road segment in our ground truth datasets, we obtained the features data from a variety of sources. To obtain the OSM feature, we queried the OSM-related APIs \cite{overpass,ors} using the coordinates of the road segments. Concerning traffic statistics features, we obtained them from government websites and iRAP websites (which pre-compiles traffic data from the government.)

% \begin{table}[h!]
% \centering
% \begin{tabular}{|p{3.0cm}|p{1.3cm}|p{1.3cm}|p{1.3cm}|p{2.9cm}|}

% \hline
% 	   \textbf{Road Safety Class} & \textbf{Croatia} &\textbf{Greece} &\textbf{Slovenia} &\textbf{The Netherlands} \\
	     
% 	  \hline
% 	High-risk road & 11.71\% & 8.39\% & 16.31\% & 8.00\% \\
%      \hline
%     Safe road & 88.29\% & 91.61\% & 83.69\% & 92.00\%\\
% 	 \hline
%     Total & 6320 & 41582 & 22873 & 55985\\
%       \hline

\begin{table}[h!]
\centering
\caption{The summary statistics for road safety class labels in four ground truth datasets. Our statistics show that two road safety classes are highly imbalanced. (HR: Croatia; GR: Greece; SL: Slovenia; NL: The Netherlands)}
\begin{tabular}{p{1.3cm}p{1.3cm}p{1.3cm}p{1.3cm}p{1.3cm}}

\hline
\rowcolor{LightGray} 	 \textbf{Class} & \textbf{HR} & \textbf{GR} & \textbf{SL} & \textbf{NL} \\
	     
	  \hline
	High-risk & 11.71\% & 8.39\% & 16.31\% & 8.00\% \\
    %  \hline
    Safe  & 88.29\% & 91.61\% & 83.69\% & 92.00\%\\
	 \hline
    Total & 6320 & 41582 & 22873 & 55985\\
      \hline
      
\end{tabular}

\label{table:stats_labels}
\end{table}

\begin{table}[h!]
\centering
\caption{The missing values rates for OSM features. Features that are not shown in the table do not have missing value problems. Our statistics show that the missing values rates differ among different countries. (HR: Croatia; GR: Greece; SL: Slovenia; NL: The Netherlands)}
\begin{tabular}{p{1.3cm}p{1.3cm}p{1.3cm}p{1.3cm}p{1.3cm}}
% \begin{tabular}{|p{2.0cm}|p{1.3cm}|p{1.3cm}|p{1.3cm}|p{2.9cm}|}

\hline
	   %\textbf{Feature} & \textbf{Croatia} &\textbf{Greece} &\textbf{Slovenia} &\textbf{The Netherlands} \\
\rowcolor{LightGray} 	 \textbf{Feature} & \textbf{HR} & \textbf{GR} & \textbf{SL} &\textbf{NL} \\
	  \hline
    oneway & 9.2\% & 23.3\% & 12.03\% & 19.7\% \\
    %  \hline
    maxspeed & 33.2\% & 62.7\% & 41.0\% & 51.4\%\\
% 	 \hline
    surface & 35.7\% & 43.1\% & 44.9\% & 50.3\%\\
    % \hline
    smoothness & 65.3\% & 78.5\% & 68.2\% & 73.5\%\\
    % \hline
    lit & 41.3\% & 47.2\% & 45.8\% & 47.6\%\\
    \hline
    average & 36.94\% & 50.96\% & 42.38\% & 48.5\% \\
      \hline
\end{tabular}
\label{table:stats_features}
\end{table}

\subsubsection{Data Characteristics and Preprocessing}

Our ground truth datasets have certain characteristics that require additional preprocessing. First, the two classes for road safety are imbalanced. Table \ref{table:stats_labels} shows that, consistently across all four datasets, only a small fraction of the road segments (8\% - 17\%) are in the high-risk class, presenting challenges for some machine learning algorithms \cite{he2009learning}. As such, we applied a data re-sampling technique, a common countermeasure for handling class-imbalance problems in applied machine learning research (e.g., \cite{dataresample_1,dataresample_2,dataresample_3}). Specifically, we applied the commonly-used Synthetic Minority Over-sampling Technique (SMOTE) \cite{smote} on our training set, which over-samples minority class by creating "synthetic" samples that fall within the proximity of real samples.

Second, as shown in Table \ref{table:stats_features}, OSM feature qualities vary among four ground truth datasets. Like many other peer-production projects such as Wikipedia, OSM does not stipulate mandatory attributes. Instead, OSM gives contributors the freedom to add as many, or as few attributes. Our analyses show that, while most of the OSM features we defined have high data completeness, a handful of features, especially features that represent road attributes, suffer from missing value problems. More importantly, analysis presented in Table \ref{table:stats_features} indicates that, for these OSM features, the specific rates of missing values are different for different countries we studied, a trend that is consistent with prior studies on OSM data quality \cite{notathome,antoniou2015measures,haklay2010many}. For example, while the rates of missing values are lower in the Croatia (avg. 37\%) dataset and the Slovenia dataset (avg. 42\%), they are considerably higher for Greece dataset (avg. 51\%) and the Netherlands dataset (avg. 49\%). To partially address the varying degree of missing data, we applied the SOFT-IMPUTE algorithm \cite{SoftImpute}, a data imputation technique that employs a soft-thresholded singular value decomposition to calculate the missing elements in the feature matrix iteratively.

\subsection{Training and Evaluation}
We trained our road safety classifier with a variety of machine learning algorithms, including both ensemble models and conventional models. With respect to ensemble models, we used an XGBoost model \cite{xgboostfirst} to perform the classification. In addition to the ensemble model, we also employed a number of traditional classifiers, including Logistic Regression, Random Forest, and Decision Tree. Our experiments revealed that XGBoost consistently outperformed other machine learning algorithms. As such, we report only the classification accuracies for XGBoost in the Results section.

To build training and testing sets, we can not use standard k-fold cross-validation due to the particular nature of our dataset. Road segments, like many other types of spatial data, are adjacent to each other. As such, training and test sets generated by standard k-fold cross-validation will be correlated, violating the independent identical distribution (i.i.d) assumption fundamental to many statistical machine learning algorithms \cite{GPSVirtual}. As such, we adopted a strategy from Song et al. \cite{song2018farsa} to maximize the independence between the samples in the training set and the samples in the test set. Specifically, after building the initial training and test sets with the standard 10-fold cross-validation, we swapped a handful of samples between training sets and test sets to ensure that samples in the test set are at least 500 meters away from the samples in the training set.

For evaluation, we report the average precision, recall, and F1 score using 10-fold cross-validation with the above spatially-aware strategy. We selected precision, recall, and F1 score as performance metrics because these metrics are more informative for evaluating performance on imbalanced data \cite{saito2015precision}. In addition, we report the performance for high-risk road class and safe road class separately to highlight the different extent of performance improvements for each class. Because this paper is the first to define and to attempt to predict road safety labels in the context of navigation systems, we cannot compare our classifier's performance with prior work. This is a relatively common scenario when applying machine learning to HCI research due to HCI's nature as a problem-defining discipline (e.g., \cite{hecht2011tweets,pang2002thumbs,lin2017problematizing}). As such, we followed the best practice to compare the performance of our classifier with that of the straightforward baseline approaches \cite{hecht2011tweets,pang2002thumbs,lin2017problematizing}. In this paper, we leveraged the most powerful baseline approach, which randomly generates predictions by respecting the class distributions in ground truth dataset.

\subsection{Results}
Results in Table \ref{table:result} show our model's ability to predict safety labels for road segments in our four country-defined ground truth datasets. Our results highlight three high-level trends. First, across the board, our road safety classifier consistently outperformed the baseline approach on all metrics. In particular, the performance increase is most significant for high-risk road class - the minority class. For example, on the Netherlands dataset, our road safety classifier that used all features can increase the F1 score on the high-risk road class from 0.07 to 0.62, a nine-fold increase. Similarly, considerable performance increase (at least five-fold increase) can be found in the results for high-risk road class in other datasets. Even for the safe road class, which already has very high baseline accuracies, our road safety classifier still consistently make smaller improvements. For example, on the Greece dataset and the Netherlands dataset on which our classifier can achieve over 0.9 baseline F1 scores for safe road class, our classifier trained on all features still further increases the F1 scores by about 0.05. Overall, these results demonstrate the clear benefits of applying our road safety classifier - for areas without existing road safety labels, applying our road safety classifiers offers predictions that are consistently better than baseline approaches. On the Croatia dataset, our classifier even reaches near-perfect predictions.

The second trend in Table \ref{table:result} is that although our classifier consistently outperformed the baseline, we notice that our classifier's absolute accuracies, especially for the high-risk road class, vary among different ground truth datasets. For example, our classifier achieves high accuracies with over 0.8 F1 scores for high-risk road class on the Croatia dataset and on the Slovenia dataset. In comparison, F1 scores for high-risk road class on the Greece dataset and the Netherlands dataset are moderate (\texttt{\~{}}0.55). One possible explanation is the different OSM feature qualities in these four countries - recall Table \ref{table:stats_features} shows that the Greece dataset and the Netherlands dataset have high rates of missing values (\texttt{\~{}}50\%) for some OSM features, while the Croatia dataset and the Slovenia dataset have lower rates of missing values (\texttt{\~{}}40\%). Although our classifier's performance on the lower-end still outperformed the baseline significantly, we suspected that the performance could further decrease when applied to regions that have even higher missing values rates. We return to this point in the discussion section.

Table \ref{table:result} shows that the parsimonious models using only OSM features perform almost as well as the models using all features. Our results show only a negligible decrease in the F1 score for most of the experiments. The fact that parsimonious models are able to perform equally well demonstrates the robustness of our approach - even in areas where local government does not provide traffic statistics, the performance of our road safety classifier will not significantly decrease.

\begin{table}[!ht]

\caption{Our road safety classifier's performance on the Croatia dataset. The performance of the baseline approach is in the bracket. (HR: Croatia; GR: Greece; SL: Slovenia; NL: The Netherlands)}
\centering
\begin{tabular}{p{0.5cm}|p{1cm}p{1.1cm}p{1.25cm}p{1.25cm}p{1.25cm}}
\hline
% \cline{2-6}
	 \rowcolor{LightGray} & \textbf{Class} & \textbf{Features} &\textbf{Precision} & \textbf{Recall} &\textbf{F1-Score}\\
	 \hline
	\multirow{4}{.5cm}{\centering HR} & \multirow{2}{1cm}{\centering High-risk}  & All  &0.95 (0.13) &0.94 (0.12) &0.94 (0.12) \\
% 	\cline{3-6}
    & &OSM &0.93 (0.13) &0.94 (0.12) &0.93 (0.12) \\ 
    \cline{2-6}
	& \multirow{2}{1cm}{\centering Safe}  &All  &0.99 (0.88) &0.99 (0.89) &0.99 (0.88) \\
% 	\cline{3-6}
    & & OSM&0.97 (0.88) &0.98 (0.89) &0.98 (0.88) \\ 
    \hline
    \hline
    
    \multirow{4}{.5cm}{\centering GR} & \multirow{2}{1cm}{\centering High-risk} & All  &0.58 (0.08) &0.45 (0.08) &0.51 (0.08) \\ 
    % \cline{3-6}
    & & OSM &0.57 (0.08) &0.44 (0.08) &0.50 (0.08) \\ 
    \cline{2-6}
	 & \multirow{2}{1cm}{\centering Safe}    &All  &0.95 (0.92) &0.97 (0.92) &0.97 (0.92) \\ 
% 	\cline{3-6}
    & & OSM &0.93 (0.92) &0.96 (0.92) &0.96 (0.92) \\
    \hline
    \hline

    \multirow{4}{.5cm}{\centering SL} & \multirow{2}{1cm}{\centering High-risk} & All  & 0.79 (0.16) &0.83 (0.16) &0.81 (0.16) \\ 
    % \cline{2-5}
    & & OSM & 0.78 (0.16) &0.81 (0.16) &0.80 (0.16) \\ 
    \cline{2-6}
	 & \multirow{2}{1cm}{\centering Safe} &All  & 0.95 (0.83) &0.95 (0.83) &0.95 (0.83)\\ 
% 	 \cline{2-5}
    & & OSM & 0.94 (0.83) &0.95 (0.83) &0.94 (0.83) \\
    \hline
    \hline
    
    \multirow{4}{.5cm}{\centering NL} & \multirow{2}{1cm}{\centering High-risk} & All  & 0.74 (0.07) &0.53 (0.08) &0.62 (0.07) \\ 
    % \cline{2-5}
    & & OSM & 0.71 (0.07) &0.44 (0.08) &0.54 (0.07) \\ 
    \cline{2-6}
	 & \multirow{2}{1cm}{\centering Safe} &All  & 0.96 (0.92) &0.97 (0.92) &0.97 (0.92) \\ 
% 	 \cline{2-5}
    & & OSM & 0.93 (0.92) &0.96 (0.92) &0.94 (0.92) \\
    \hline
    
\end{tabular}
\label{table:result}
\end{table}

\section{Discussion}
Our results demonstrate that using straightforward machine learning algorithms and publicly available geographic data as features, we can successfully predict road safety labels with accuracies that are significantly higher than the baseline approach. In the following, we first discuss the factors to consider when extending our road safety classifier to regions beyond the areas studied in this paper. With an eye towards future work, we then discuss various technology designs through which our road safety predictions could help personal navigation technologies reduce the risks of dangerous routes.

\subsection{Extending our approach to other regions}
Our results show that, while the performance of our classifier consistently beat the baseline, its performance in a region might be affected by the OSM feature quality in that region. 

As such, when applying our road classifier to other regions, one important factor to consider is the quality of OSM features in this region. Researchers have found that OSM data quality is usually better in urban/developed areas than in rural/undeveloped areas \cite{notathome,proprietarygeodata,crowsourcing}. As such, our road safety classifier might have higher performances for more developed regions such as Europe and have lower performances for less developed regions, such as South Asia and Africa. Similarly, urban areas, which have higher OSM qualities, might observe higher performance than the rural areas. As such, future work that extends our classifier to other regions should examine the specific OSM data quality, especially the rates of the missing value, before applying our approach. Moreover, in order to improve the accuracy in the areas with low OSM data quality, future work should explore additional features beyond OSM and traffic statistics or more advanced machine learning models.

\subsection{Integrating Road Safety Predictions into Navigation Systems}
Our work can produce large-scale reliable road safety predictions which navigation systems can directly leverage. These road safety predictions present many opportunities for designing new algorithms and user interactions for protecting drivers' safety. Below, we briefly mention the opportunities for new algorithm designs and then focus our attention on new user interactions that ensure driver safety.

For algorithm innovation, one clear opportunity is building new routing algorithms that prioritize road safety. As a starting point, one may simply adopt the existing algorithms designed for generating safe routes that pass through low-crime areas (e.g., \cite{kim2014socroutes,fu2014treads,shah2011crowdsafe}). For example, the crime-index score in the cost functions of these safe routing algorithms can be replaced by an aggregated road risk score of the candidate route. Alternatively, instead of adding waypoints in low-crime areas, these safe routing algorithms can add waypoints that detour through a safe road.

However, even with new routing algorithms that prioritize road safety, drivers might still unavoidably travel on high-risk roads given their time constraints or their choices of destinations. In these scenarios, road safety awareness-enabled user interactions serve as the critical last layer of protection. These essential user interactions can happen in two phases - during route recommendation and while driving. During route recommendations, navigation systems can put users in control of the trade-off between road safety and other parameters. To begin with, as a preference setting, navigation systems can explicitly ask for the maximum distance or portion of high-risk roads that users are willing to tolerate within a route and incorporate this constraint when running the routing algorithms. In addition, navigation systems can explicitly communicate the trade-off between time and route safety among alternative routes through routes overview. For example, in addition to informing users of the distance and time of the candidate routes, navigation systems can highlight the distance or the portion of high-risk road segments through textual or visual communications.

During the trip, when passing through high-risk roads, drivers must focus their entire attention on the road \cite{horberry2006driver}. As such, HCI researchers have called for navigation systems that can timely alter drivers for abnormal road environments (e.g., \cite{casner2016challenges}). With our road safety predictions, navigation systems will know precisely where the high-risk road segments are. Given this knowledge, navigation systems can generate audio and visual alerts when drivers are approaching the dangerous road segments to call drivers' attention to the road. With cars becoming increasingly automated and drivers delegating more responsibility to the technology, navigation systems with road safety awareness will be even more vital to ensure user safety.

%Conclusion
\section{Conclusion}
A dangerous assumption adopted by existing navigation systems is that all roads are equally safe. Navigation systems with this wrong assumption have routed drivers to dangerous routes, threatening driver safety. In this paper, we sought to address this risk by adding road safety awareness to the navigation systems. Specifically, we defined road safety in the context of the navigation systems and proposed a road safety classifier to automatically predict the road safety for many roads using out-of-the-box machine learning algorithms and diverse features from public geographic data. We demonstrate we can predict the road safety labels with accuracies that are significantly higher than baseline approaches. Finally, we discussed the factors to consider when extending our road safety classifier to other regions and the novel safety designs on navigation systems enabled by our road safety predictions.

\bibliographystyle{ACM-Reference-Format}
\bibliography{sample-base}
\end{document}